# Having our 'omic' cake and eating it too: Evaluating User Response to using Blockchain Technology for Private & Secure Health Data Management and Sharing


**Victoria L. Lemieux, PhD** (corresponding author)

University of British Columbia, Rm 488, 1961 East Mall, Vancouver, BC, V6T 1Z1.
Email:vlemieux@mail.ubc.ca

**Darra Hofman, JD, MSLS**

University of British Columbia

**Hoda Hamouda, MDes**

University of British Columbia

**Danielle Batista, MIS**

University of British Columbia

**Ravneet Kaur, MEng**

University of British Columbia

**Wen Pan, BComm**

University of British Columbia

**Ian Costanzo, BASc**

Anon Solutions

**Dean Regier, PhD**

BC Cancer
University of British Columbia

**Samantha Pollard, PhD**

BC Cancer

**Deirdre Weymann, MA**

BC Cancer

**Rob Fraser, PhD**

Molecular You




**This paper reports on the development and evaluation of a prototype blockchain solution for private and secure individual "omics" health data management and sharing. This solution is one output of a multidisciplinary project investigating the social, data and technical issues surrounding application of blockchain technology in the context of personalized healthcare research.  The project studies potential ethical, legal, social and cognitive constraints of self-sovereign healthcare data management and sharing, and whether such constraints can be addressed through careful user interface design of a blockchain solution.**

## Introduction

There is a news story almost every day about how individuals' personal data are being harvested, shared with and used by third parties without their consent and in ways that have real potential to cause harm. The result is an erosion of user trust and a reluctance to use services that gather sensitive information (Edelman, 2019). This remains true for a significant percentage of individuals even if they could greatly benefit from receiving a personalized health service that they can use to understand their health risks and maintain or improve their overall health (Betts and Korenda, 2018; Shabani, Bezuidenhout, and Borry, 2014). Individuals' reluctance may stem from uncertainty about how health data services will store and use their data over time (Sanderson et al. 2016; Shabani, Bezuidenhout and Borry, 2014). Recent revelations about how Facebook, 23&Me, and other platforms use individuals' sensitive personal data validates concerns that consumers' data may be shared with third parties without their informed consent (see, e.g., Rosenberg, 2018; Geggel, 2018).





This paper reports on an ongoing multidisciplinary research project that responds to concerns about health data privacy and sharing, in particular, through investigation of the potential application of blockchain technology. Blockchains are distributed ledgers in which confirmed and validated blocks are organized in an append-only chain using cryptographic links in support of the goal of decentralized, autonomous trust (InterPARES 2017). Some argue that blockchain technology can be used to provide individuals with control over their own data (i.e., "self-sovereignty" [Allen, 2016]) to prevent the kind of "databuse" (Wittes, 2011) that has reduced individuals' trust in sharing their data. On the other hand, blockchain technology is an emerging technology that, thus far, has proven difficult for all but experts to grasp. Research has shown that blockchain has a usability problem (Kromholz, et al., 2016; Eskandari et al., 2018). Given this reality, it is fair to ask if data self-sovereignty enabled by blockchain technology is really the solution to protecting individual's health data.  While there have been many studies of using blockchain technology in healthcare (Zyskind, Nathan and Pentland, 2015a & b; Peterson et al, 2016; Mackay and Nayyar, 2017; Xia et al., 2017; Dubovitskaya et al., 2017; Ferdous et al., 2017; Brogan, Baskaran and Ramachandran, 2018; Griggs et al., 2108; Kaur et al., 2018; Zhang et al., 2018; Zhou, Wang and Sun, 2018), we actually know very little about how individuals might actually respond to controlling their own data using blockchain technology. Thus, to shed greater light on the use of blockchain technology to enable data self-sovereignty from the end user's perspective, we developed a technical artefact – the self-sovereign health data management blockchain solution design. We then used the artefact to stimulate a conversation with focus group participants to learn more about how individuals would respond to being given control over their own health data and using a blockchain solution to manage and share their data.





## Background Literature

*Advancing personalized medicine: Why both data sharing and data privacy matter*

Omics science, including genomics, proteomics, exposomics, phenomics, microbiomics and metabolomics (Horgan and Kenny, 2011), provides insights into health at a molecular level never before possible and has the potential to radically alter healthcare. Omic science establishes a sophisticated, systemic understanding of the "complex, longitudinal, and dynamic nature of biological networks (and their fluctuations in response to social/environment exposures) that fundamentally govern human health and disease" (Holmes et al., 2010, 327). Indeed, Bencharit (2012) asserts that "the new era of omics studies . . . may lead to a true clinical application of personalized medicine."

The undeniable social good that omics could do is not without challenges and risks, however. Privacy for participants in research and in clinical applications is a major concern because "[b]y nature, the genome encodes a sensitive yet heritable signature of an individual that is marked by genetic variation reflecting one's ancestry and disclosing one's susceptibility to health and diseases" (Shi and Wu, 2017, 61). Both Canada and the United States have passed genetic non-discrimination acts (e.g., Genetic Non-Discrimination Act, S.C.2017, c.3; Genetic Information Non-Discrimination Act, 29 USC §216(e), 29 USC §1132) in light of the potential medical, professional, legal and social consequences that individuals might face should their genomic information be disclosed. Other omic information also has the same potential for abuse. Given the very grave potential consequences of unauthorized disclosure of omic data, protecting the privacy of individuals is of paramount importance.

Family privacy is also a concern, since omics science extends not just to the individual, but to their family as well (Shi and Wu, 2017, 61). After all, genes are heritable – breaching the genetics of one





individual may easily reveal private information about those who share that individual's genes. "Clinical genetics guidelines [in the United Kingdom] conceptualise genetic information as confidential to families, not individuals" (Dheensa, Fenwick and Lucassen, 2017, 1).

Beyond consideration of the consequences of privacy breaches, however, lies a deeper reason to ensure that individuals' privacy is protected. In a world in which we increasingly live online, we are our data and are data are us (Cheney-Lippold, 2018). The philosopher and information ethicist, Luciano Floridi views consequentialist ethical frames of reference that focus on judging actions as moral or not based on their outcomes as insufficient (Floridi, 1999). He writes that, "Typically, privacy and confidentiality are treated as problems concerning S' ownership of some information, the information being somehow embarrassing, shameful, ominous, threatening, unpopular or harmful for S' life and well-being, yet this is very misleading, for the nature of the information in question is quite irrelevant. It is when the information is as innocuous as one may wish it to be that the question of privacy acquires its clearest value. The husband, who reads the diary of his wife without her permission and finds in it only memories of their love, has still acted wrongly. The source of the wrongness is not the consequences, nor any general maxim concerning personal privacy, but a lack of care and respect for the individual, who is also her information." (Floridi, 1999, p. 53). Thus, we see in Floridi an approach that views an individuals' data as equivalent to the individual themselves, which suggests that to abuse a person's data is tantamount to an assault of their physical being.

Simply locking data away, then, is a poor solution. "Sharing genetic findings is vital for accelerating the pace of biomedical discoveries and for fully realizing the promises of the genetic revolution" (Erlich and





Narayanan, 2014, 409). Thus, if omic research is to be utilized to its full potential, solutions must be found to protect privacy while still permitting data sharing and usage.

*Blockchain Technology: A possible solution to private & secure data sharing*

Blockchain's design and networked, distributed, autonomous, and global operation establish it as a disruptive technology with social, political, and economic implications that far exceed those of other emerging technologies with many potential applications (Economist, 2015; Casey & Vigna, 2018). One of the key applications identified has been in connection with decentralized management of data and privacy.

Swan (2015) notes that, by managing electronic medical records in the blockchain, they "could be analyzed but remain private, with an embedded economic layer to compensate data contribution and use." She also envisions "a standardized secure mechanism for digitizing health data into a health data commons" where patients could consent to making their health data available for research use in exchange for cryptocurrency. Benchoufi and Ravaud advocate for blockchain to address "reproducibility, data sharing, personal data privacy concerns and patient enrolment" (2017, 335), and emphasize "the transparency of the Blockchain ledger – owned by no one, publicly writable by anyone […] users do not need any third party to trust the system" (2017, 338). And, Gropper (2016) proposes the application of a decentralized identity management solution within the healthcare sector.

 With the level of trust that it can enforce, blockchain also could be considered a path through the complexities of user consent. Meaningful consent is critical if health data is to be used both ethically and legally with "[C]onsent [being] a cornerstone of both biomedical research ethics and data protection





law" (Thorogood and Zawati, 2015, 693). A number of studies have aimed to apply blockchain technology to giving individuals direct control over access to their medical records and consenting to secondary use of their health data for research purposes. Ekblaw et al (2016), Ivan (2016), Broderson et al (2016), Li et al (2017), Linn and Koo (2016), and Dagher et al (2018) discuss blockchain-based medical records systems that incorporate user-defined permissioning while still storing patient records in a provider's existing systems. Yue et al (2017) propose the *Healthcare Data Gateway* application to allow users to control their own health data and permission its use for research purposes. Zhang et al (2018) present a decentralized application for patient-defined access to structured pieces of their health data record and Patel (2018) discusses a blockchain-based framework for medical image sharing that allows for patient-defined access permissions. Finally, Hofman et al (2018) discuss a blockchain prototype for managing user consent in the use of clinical data for precision health research.

While blockchain could be a solution to some of the challenges of securing and protecting patients' health data (Engelhardt, 2017), giving patients greater control over their data using this technology, it is not without its challenges (Gordon and Catalini, 2018). The cryptography and networking involved in blockchain technology can make it difficult for even IT specialists to understand, let alone users (Lunggren, 2019).  Many patients already have difficulty navigating the healthcare system, which raises questions about whether placing the added burden upon them of managing their own healthcare records, and associated consents to access and use of the data within these records, will truly generate a net positive effect (Gordon and Catalini, 2018). Omic data is particularly challenging in terms of meaningful, informed consent. Indeed, omic data represents an extreme form of "the transparency paradox […] If notice (in the form of a privacy policy) finely details every flow, condition, qualification, and exception, we know that it is unlikely to be understood, let alone read. […] An abbreviated, plain-





language policy would be quick and easy to read, but it is the hidden details that carry the significance"

(Nissenbaum, 2010, 36). After all, omic research techniques – and therefore research purposes –

advance quickly, making it challenging to explain the purpose, risks, and benefits of studies in an

accessible way. Indeed, it is difficult to even predict "all the informational benefits and risks of research

with complex genomic information" (Thorogood and Zawati, 2015, 694). Moreover, some bioethicists

also worry about the possibility of coercion if patients are financially incentivized to share their personal

health data (Gammon, 2018). A blockchain solution can give users greater control over access to their

health records and consent to use of their health data, but will they be able to navigate both the

complexity of consent in addition to a novel technology?  Searching for answers matters.

## Methodology

We followed a multi-method, two-stage methodology to find out more about the potential of blockchain

technology to be used to protect the privacy of individuals' personal health data and enable secure data

sharing without introducing cognitive and other barriers that might prevent users from understanding

and effectively navigating the such systems.

*Stage One: Designing a Self-Sovereign Health Data Management Solution*

In the first stage, we set out to design a technical artefact, in the form of a blockchain solution that

fundamentally respects users' right to privacy and provides them with the same level of choice and

control over the sharing of their data as they would expect over the sharing of their bodies, with a view

to exploring our research question. We decided that blockchain protocols that came closest to our vision

were those that supported self-sovereign identity . Self-sovereign identity (SSI), a variant of

decentralized digital identity, leverages the affordances of blockchain technology to increase users'





control of their identities in the digital world (Allen, 2016). It implies that individuals' identities and the data associated with them are neither bestowed, revocable nor owned by any authority save for the individual herself. Christopher Allen writes that "[s]elf-sovereign identity is the next step beyond user-centric identity […] the user must be central to the administration of identity [with] true user control of that digital identity, creating user autonomy: (Allen, 2016) Kaliya Young and Heather Vescent (2018) explain that "Self Sovereign Identity is a new technology layer that enables individuals and organizations to assert their own identity." Tobin and Reed (2017) describe Self-Sovereign Identity as "the result of trying to satisfy three basic requirements: 1. Security - the identity information must be protected from unintentional disclosure; 2. Control - the identity owner must be in control of who can see and access their data and for what purposes (see Figure 1); and 3. Portability - the user must be able to use their identity data wherever they want and not be tied into a single provider. Tim Bouma (2019) argues that in "the old (centralized and federated) models the locus of control was between the other parties that could make decisions about me, whether I was in the picture or not." The basic tenets of SSI can be summarized at a high-level as follows: 1) every individual human being is the original source of their own identity; 2) identity is not an administrative mechanism for others to control; and 3) each individual is the root of their own identity and central to its administration (IBM, 2018). This approach differs markedly from Privacy by Design (Cavoukian, 2011) and Global Alliance for Genetic Health (GA4GH)'s Framework for Responsible Sharing of Genomic and Health-Related Data (GA4GH, 2016), wherein data stewards, research ethics boards, and researchers still make decisions about a data subject's data. By contrast, with self-sovereign identity the locus of ownership and control of decision-making shifts to the individual.





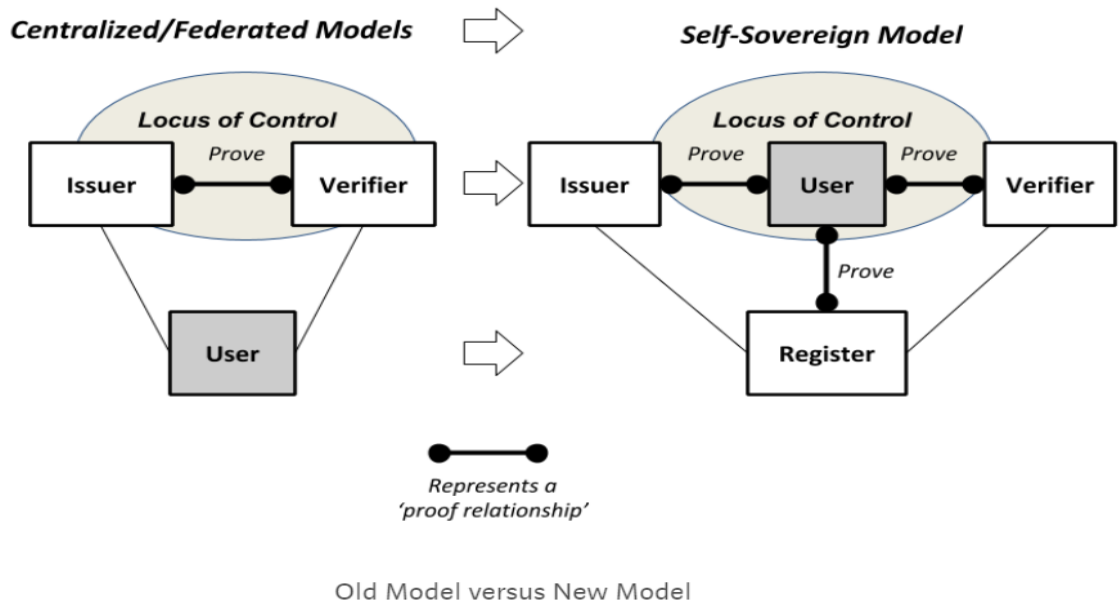

Figure 1: Self-Sovereign Identity Locus of Control (Bouma, 2019).

Having decided upon an SSI-based solution design, we created a design artefact using prototyping and agile software development. The agile approach draws upon a group-based, collaborative software development methodology that uses iterative, highly context sensitive requirements for identification, design, implementation, and evaluation. Agile development typically involves short, intense sprints wherein cross-functional teams gather in "scrums" to identify requirements, develop code, and evaluate the functionality of a proof-of-concept software application (Agile Alliance, 2013).

Given the focus of the solution design on shifting the locus of control, custody and decision-making about health data to users of the solution, we also employed user-centered design (UCD) as the general methodological approach to the design and implementation of our prototype. UCD methodology is also widely used when designing health care services (LeRouge & Wickramasinghe, 2013; Xie & Carayon,





2015) UCD ensures the involvement of users and the inclusion of their perspectives in the research, development and assessment phases of a design (Ghulam Sarwar Shah and Robinson, 2006).

The architecture of resulting technical artefact, which was developed on Hyperledger Indy (HLI), is shown in Figure 2.  HLI is comprised of four basic components: 1) verifiable claims, 2) peer-to-peer agents, 3) decentralized identifiers, and 4) a distributed ledger. Verifiable claims are a "machine-readable statement made by an entity that is cryptographically authentic (non-repudiable)" (W3C, 2019). Verifiable claims are made when a "holder" agent sends a cryptographic credential – provable digitally signed data - received from an "issuer" agent to another "verifier" agent across a peer-to-peer connection which the receiving verifier agent is able to cryptographically prove is authentic (see Figure 4). Each interacting agent has a decentralized identifier - a new type of identifier for verifiable, "self-sovereign" digital identity that is independent from any centralized registry, identity provider, or certificate authority (W3C, 2019; DIF, 2019) - used to facilitate communication in each pairwise connection. HL Indy employs privacy-enhancing techniques, such as selective disclosure and zero-knowledge proofs – a cryptographic technique allowing an agent to prove that they know something, such as a password, without revealing what they know. A distributed ledger – a ledger that is shared across a set of nodes (i.e., network endpoints) and synchronized between the nodes using a consensus mechanism (e.g., Practical Byzantine Fault Tolerance) – is used to store Public DIDs (e.g., of issuer agents), data schemas for credentials, credential definitions, and revocation registries – that enable the cryptographic verification of claims.





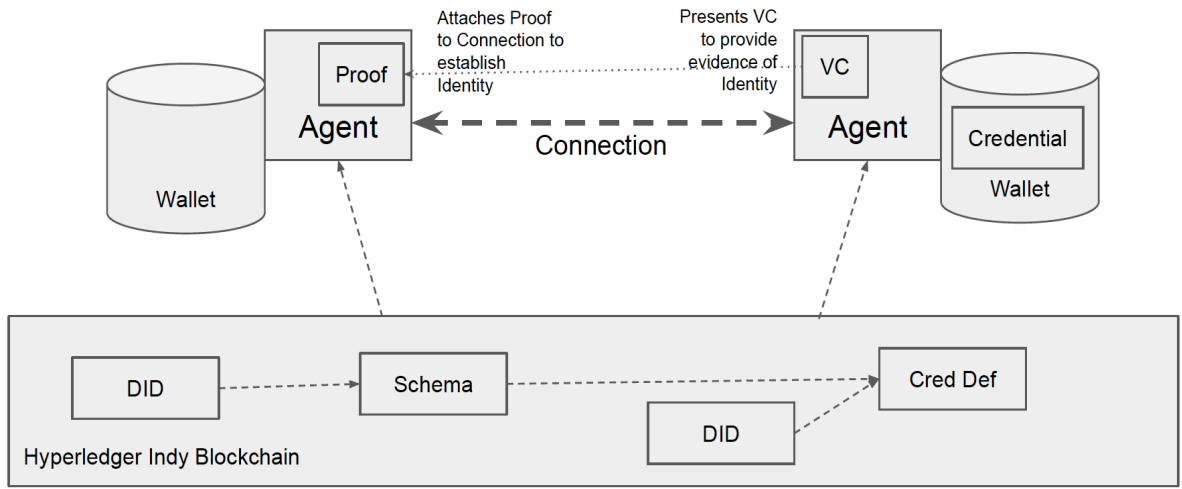

Figure 4: High Level Solution Architecture (Source: Anon Solutions).

Our specific solution design incorporated four main actors: 1) MYco, which is an issuer of individuals' health credentials; 2) Ethics Review Boards (ERB), which issue ethics credentials to researchers so that individuals can verify that their data will be handled properly when shared; 3) Researchers, who apply to the ERB to conduct research projects and market these to data owners; and 4) Data Owners, MYco clients who hold health credentials issued by MYco that are shared with researchers with the data owner's consent. The solution was designed to support a number of steps in the process of providing individuals with control of their health data and enabling privacy-preserving and secure data sharing in support of personalized health research.

The first step in the process of data owners sharing their data and receiving rewards for their contributions occurs when they request cryptographic credentials for each of their biomarkers from MYCo. MYco then issues these credentials to the personal health wallet (agent) of each individual data owner (MYco client).





 The solution also supports researchers' application to an Ethics Review Board (ERB) for ethics certificates to conduct their research, and once approved, the sending of the ethics certificate in the form of a cryptographic credential to the wallet (agent) of the applicant/researcher. Once they have ethics approval, researchers are then able to use the solution to advertise their research projects to data owners.

When a data owner notices a research project in which s/he would like to participate, this initiates a "handshake" process using a peer-to-peer connection in which data owners verify that the research project has the necessary ethics approval and researchers verify that data owners meet their study criteria and consent to the sharing of their data. The process completes when data owners share the specific health data (e.g., biomarkers) needed for the study and researchers send data owners a reward for their participation in the study in the form of a cryptographic credential. Figure 3 represents a visual overview of the "handshake" process that takes place between data owners and researchers. A key feature of the entire handshake process is its alignment with the motivating theoretical and design principles; that is, the solution is designed to ensure that the identity of data owners is never revealed to researchers, no personal health information is ever recorded or stored on the blockchain to prevent conflicts with privacy laws and reduce the potential for privacy breaches, and data owners remain in control of their personal health information at all times, revealing only as much information as they feel comfortable with given their assessment of the risk-benefits of the transaction.





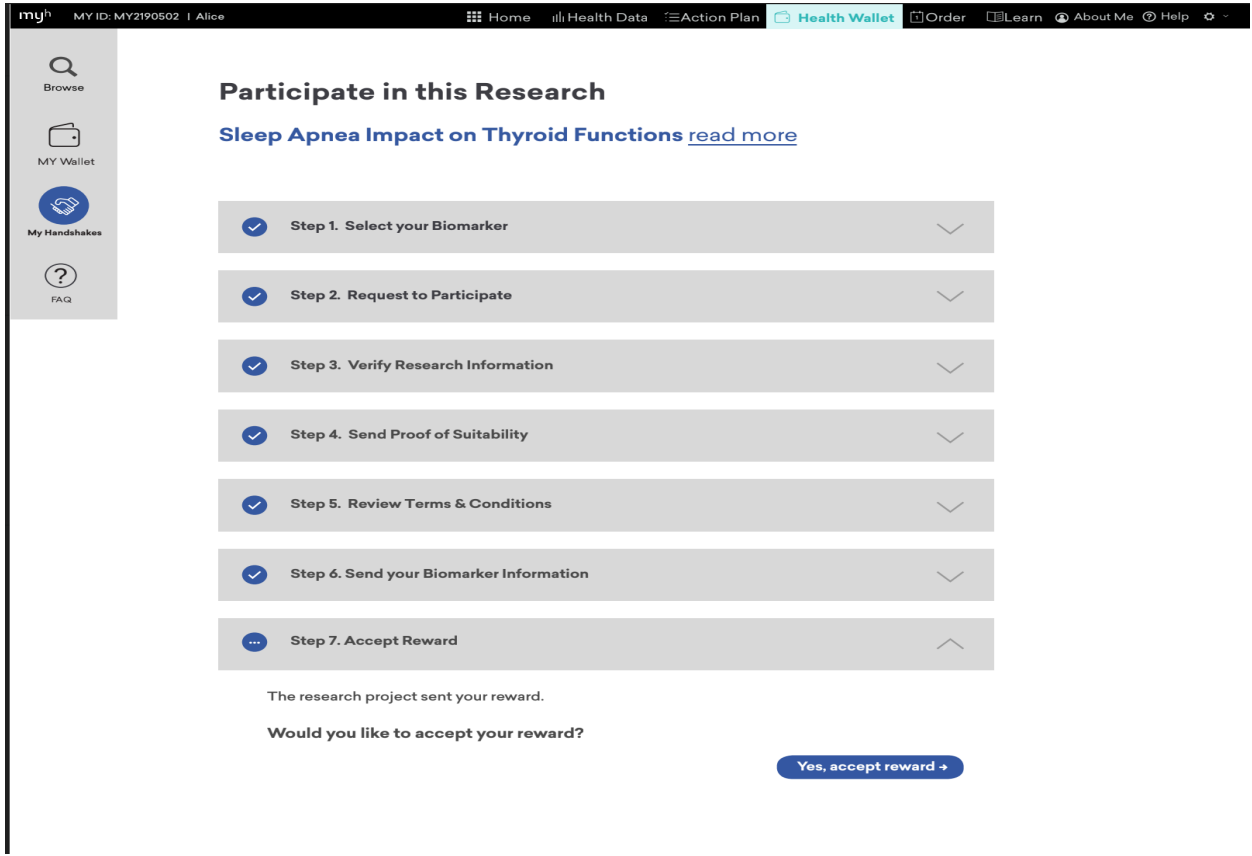

Figure 3: Wireframe showing Handshake Process.

*Stage Two: Focus Group Evaluation of Blockchain Solution Design*

*Qualitative data analysis*

As described in the previous section, validation of the solution design was done throughout the process of designing and implementing a prototype, with final evaluation of the prototype relying upon data gathered from three focus groups. Focus groups are suitable for exploring the attitudes towards new





phenomena such as blockchain, as the relatively open-ended discussions can sensitize researchers to unrealized issues and increase the comprehensiveness of large-scale surveys conducted afterward (Morgan, 2005). . In total, 26 individuals participated in our study, with eight in the first focus group, eight in the second group and ten in the third group.. The focus groups were comprised of individuals aged 25-60 years old recruited from an online graduate student group. Among all the participants, three participants recently finished their advanced degrees (Master/PhD), and the rest are all enrolled in a Master or PhD program. Five participants have an education background in information management or archival science, and eight participants are enrolled in graduate programs in the medical field. All the participants have been  patients at some point in life.

During the focus group, participants were primed with a presentation that contained information about the following topics: consent, management, privacy of personal health data and blockchain technology. Then they were shown wireframes of the user interface of the prototype solution and asked a set of semi-structured questions relating to their understanding of blockchain technology, the views of data privacy and sharing, and their thoughts on the user interface. The focus groups were audio recorded with participant consent. The audio recordings were transcribed verbatim and then the recordings were destroyed. Transcriptions were pseudonymized and coded for analysis using *NVIVO* qualitative analysis software. The research team read the participants' responses and extracted 6 main codes as shown in table 1.





Table 1: Analytic codes extracted from focus group participant statements.

| CODE | DESCRIPTION |
| --- | --- |
| Compensation and rewards | Types of rewards related to data sharing and the impacts of having them. |
| Ethics | Ethical issues and concerns in relation to health data sharing, use and control. |
| Health data | Discussions about health data. |
| Access | Questions and answers related to centralized and decentralized access, equity, difficulties. |
| Control | Discussion about the relevance of personal data control, data expiration, and ownership. |
| Sharing | Health data sharing with or without consent, who to share with and benefits and risks of sharing (not including ethical issues and concerns). |
| Privacy | Discussion about privacy issues, ownership and anonymization. |
| Systems design | The usability and design of the platform; suggestions for improvement. |
| Trustworthiness, security and comfort | Discussion about the trustworthiness, security and level of comfort with using decentralized system. |

**Findings**

Participants' responses flag a number of unresolved challenges to the adoption of blockchains as solutions for private and secure data sharing in healthcare, as well as specific areas for improvement of our specific solution design. The following sections provide a high-level summary of participants' feedback.

Focus group participants were generally aware of the challenges of data sharing across healthcare providers. For example, they noted that hospitals could not easily share with one another and that





moving across jurisdictions often meant losing access to their health records. They also were aware of cases when very sensitive health information had been inadvertently exposed.

Individuals saw value in using a blockchain-based solution as a means to support privacy-preserving data sharing. However, some individuals expressed reluctance to use such a platform until it has been thoroughly tested and more widely adopted. Areas of ongoing concern included who they would be sharing with and for what purpose. Generally, participants expressed willingness to consent to having university researchers use their data, or to share it with government agencies in the event of a public health crisis but were reluctant to share with pharmaceutical companies or insurers for fear of being discriminated against. This highlights the importance of designing upfront information about the type of organization requesting access and a clear explanation of their reason for wanting to use individuals' health data. Individuals also wanted assurances that researchers or other users of their data would not be able to reuse data for another purpose without their consent or assemble data about them from disparate sources to create a health profile about them (a "mosaic effect" [Wittes, 2011]). Participants were not universally hesitant to engage with a more experimental platform; as one focus group participant put it: ". . . someone has to start, right? There would be falls and all that and there would be corrections, I'm willing to be on the beta."

One cognitive constraint leading to possible lack of trust was in connection with the way that the cryptographic proofs operated. Focus group participants expressed a lack of understanding and need for more transparency about the manner in which cryptography protected privacy and validated claims, with one participant referring to the proofs as a "black box". This suggests a need for informational tools and techniques, such as decision aids that could support participants' choices to engage with the platform.





(Williams et. al, 2014) or algorithmic transparency.  Unlike in artificial intelligence (AI) solutions where solution designers have often resisted requests to reveal their algorithms in order to protect their interests (Diakopolous, 2016), there is a longstanding practice of algorithmic transparency in cryptography. Kerchoff's Principle, one of the guiding axioms of cybersecurity solution design, specifies that a cryptosystem should be secure even if everything about the system, except the private key, is public knowledge (Stewart, Tittel & Chapple, 2008).  Thus, cybersecurity solution designers have much stronger incentives for revealing their cryptographic algorithms than do AI researchers, suggesting that this cognitive barrier can be overcome.

Focus group participants generally liked the idea of having greater control and custody of their personal data, though one participant did express concern: "My first impression was 'crap, now I have to keep track of it all'." Another participant said they would share the power of control and consent with immediate family members in case anything happened to them. Universally, participants did not want to bear the risk, typical of current blockchain solutions, of losing access to their data if they lost their private cryptographic key.  They were all willing to give up some self-sovereignty for the ability to have a way to regain access.

In terms of usability of a decentralized cryptosystem, individuals expressed a number of concerns. In particular, some participants identified the risk of exclusion of non-tech savvy and older users. However, another participant in an older age demographic noted: ". . .actually today older people have more access to smartphones then they have had in the last 5 or 10 years." Another noted, "I think it will come to a stage that it will be much easier to use for older people." Participants also expressed concern about the understandability of consent terms and conditions, pointing to the fact that these statements can be very





complex and difficult to interpret, which is consistent with the findings of previous studies.  They requested that terms and conditions be presented in understandable language upfront in the handshake process, not at step five as in the technical prototype they were shown.

In relation to the offering of a reward, most individuals felt comfortable with this idea but did express some concern about potential effects in relation to the scale and granularity of data being shared and the use to which the data would be put. For example, one study participant wondered: "would that become a barrier for researchers who didn't have that kind of [money], that a company has to compensate people, and how would that affect the landscape of information sharing?" Another said, "I would also worry that the outcomes would then be skewed because if you're putting forth opportunities for compensation, then especially if you're talking $50 or less, who are you attracting? Are you really attracting a broad enough range of people that have data that's applicable to whatever the study is, so I don't like that idea." As a result, participants generally expressed a preference for smaller rewards functioning more as honoraria rather than market-based compensation. Others wanted to know more about the form a reward would take. For example, if provided in the form of a gift card, participants wondered if, they could be traced back to the research study.  As a result, some participants expressed a preference for the reward in the form of cryptocurrency, like Bitcoin, or even food. Overall, users noted that they have higher levels of trust in the process knowing that a research ethics board has reviewed the study design, including the issue of compensation, even if that meant the platform was not fully decentralized.

## Conclusion

No one solution can solve the challenges of protecting participant's privacy – of respecting their autonomy and dignity – in complex, revealing areas such as omic science. However, blockchain





technology could solve a number of the technical and social limitations of our current systems for onboarding participants and collecting, storing, and disseminating data. As Dove, Ozdemir, and Joly (2012, 439) remind us, "open innovation models, such as open access, open source, expert sourcing, and patent pools" are one of the primary means of "overcoming the 'transfer problem' in omics research that continues to hinder the full realization of concrete applications for human health" (2012, 439). One of the major hindrances to the full embrace of open innovation in omic science is the very real danger to patient privacy breach should their data be subject to unauthorized access or disclosure. Blockchain technology could let us have our omic cake and eat it too, by permitting the data to be studied while remaining private. Nevertheless, the above evaluation flags a number of ongoing areas of concern and future research challenges.

## ACKNOWLEDGMENTS

To be added.

## REFERENCES


Agile Alliance. (2013). *What Is Agile Software Development?* Retrieved from Https://www.agilealliance.org/agile101/

Allen, C. (2016, April 25). The Path to Self-Sovereign Identity. Retrieved from Life with Alacrity (blog), website: http://www.lifewithalacrity.com/2016/04/the-path-to-self-sovereign-identity.html

Bencharit, S. (2012). Progresses and challenges of omics studies and their impacts in personalized medicine. *Journal of Pharmacogenomics & Pharmacoproteomics*, *3*(1), 10001e105.

Benchoufi, M., & Ravaud, P. (2017). Blockchain technology for improving clinical research quality. *Trials*, *18*(1), 335.






Betts, D., & Korenda, L. (2018, September 25). *Inside the patient journey: Three key touch points for consumer engagement strategies*. Retrieved from https://www2.deloitte.com/insights/us/en/industry/health-care/patient-engagement-health-care-consumer-survey.html

Bouma, T. (2019, March 2). Self-Sovereign Identity: Shifting the Locus of Control. Retrieved from Medium website: https://medium.com/@trbouma/self-sovereign-identity-shifting-the-locus-of-control-10da1c8757ad

Broderson, C., Kalis, B., Leong, C., Mitchell, E., Pupo, E., & Truscott, A. (2016). *Blockchain: Securing a New Health Interoperability Experience*. Retrieved from https://www.healthit.gov/sites/default/files/2-49-accenture_onc_block-chain_challenge_response_august8_final.pdf

Brogan, J., Baskaran, I., & Ramachandran, N. (2018). Authenticating Health Activity Data Using Distributed Ledger Technologies. *Computational and Structural Biotechnology Journal*, *16*, 257–266.

Casey, M. J., & Vigna, P. (2018). *The Truth Machine: The Blockchain and the Future of Everything*. New York, NY: St. Marten's Press.

Cavoukian, A. (2011). *Privacy by Design: The 7 Foundational Principles*. Retrieved from https://www.ipc.on.ca/wp-content/uploads/Resources/7foundationalprinciples.pdf.

Cheney-Lippold, J. (2018). *We are data: Algorithms and the making of our digital selves*. NYU Press.

Dagher, G. G., Mohler, J., Milojkovic, M., & Marella, P. B. (2018). Ancile: Privacy-preserving framework for access control and interoperability of electronic health records using blockchain






technology  Dagher, G. G., Mohler, J., Milojkovic, M. and Marella, P. B. *Sustainable Cities and Society*, *39*, 283–297.

Dheensa, S., Fenwick, A., & Lucassen, A. (2017). Approaching confidentiality at a familial level in genomic medicine: A focus group study with healthcare professionals. *BMJ Open*, *7*(2), e012443.

(DIF) Decentralized Identity Foundation. (2019). *Together We're Building a New Identity Ecosystem*. Retrieved from https://identity.foundation/.

Diskopolous, N. (2016). Accountability in algorithmic decision making. *Communications of the ACM*, *59*(2), 56–62.

Dove, E. S., Ozdemir, V., & Joly, Y. (2012). Harnessing omics sciences, population databases, and open innovation models for theranostics-guided drug discovery and development: Omics sciences, databases, and open innovation. *Drug Development Research*, *73*(7), 439–436.

Dubovitskaya, A., Xu, Z., Ryu, S., Schumacher, M., & Wang, F. (2017). Dubovitskaya, A., Xu, Z., Ryu, S., Schumacher, M., and Wang, F. (2017). Secure and trustable electronic medical records sharing using blockchain. In AMIA Annual Symposium Proceedings (Vol. 2017, p. 650). American Medical Informatics Association. *AMIA Annual Symposium Proceedings*, *2017*, 650. American Medical Informatics Association.

Econonist. (2015, October 31). *Blockchains: The great chain of being sure about things*. Retrieved from https://tinyurl.com/y76dovsm.

Edelman. (2019). *Edelman Trust Barometer. Annual global survey*. Retrieved from https://www.edelman.com/trust-barometer/

Ekblaw, A., Azaria, A., Vieira, T., & Lippman, A. (2016). *MedRec: Medical Data Management on the Blockchain*. Retrieved from http://dci.mit.edu/assets/papers/eckblaw.pdf







Engelhardt, M. A. (2017). Hitching healthcare to the chain: An introduction to blockchain technology in the healthcare sector. *Technology Innovation Management Review*, *7*(10), 22–34.

Eskandari, S., Clark, J., Barrera, D., & Stobert, E. (2018). A first look at the usability of bitcoin key management. *arXiv preprint arXiv:1802.04351*.

Ferdous, M. S., Margheri, A., Paci, F., Yang, M., & Sassone, V. (2017). Ferdous, M. S., Margheri, A., Paci, F., Yang, M., and Sassone, V. (2017, June). Decentralised runtime monitoring for access control systems in cloud federations. In Distributed Computing Systems (ICDCS), 2017 IEEE 37th International Conference on (pp. 2632-2633). IEEE. *Proceedings of the 37th International Conference on Distributed Computing Systems (ICDCS)*, 2632–2633. IEEE.

Floridi, L. (1999). Information ethics: On the philosophical foundation of computer ethics. *Ethics and Information Technology*, *1*(1), 33–52.

(GA4GH) Global Alliance for Genomics & Health. (2016, March 15). *Data Sharing Lexicon*. Retrieved from https://www.ga4gh.org/wp-content/uploads/GA4GH_Data_Sharing_Lexicon_Mar15.pdf

Gammon, K. (2018). Experimenting with blockchain: Can one technology boost both data integrity and patients' pocketbooks? *Nature Medicine*, *24*(4), Online.

Geggel, L. (2018, July 28). *23 and Me is Sharing Genetic Data with Drug Giant*. Retrieved from https://www.scientificamerican.com/article/23andme-is-sharing-genetic-data-with-drug-giant/

Ghulaum Sarwar Shah, S., & Robinson, I. (2006). User involvement in healthcare technology development and assessment: Structured literature review. *International Journal of Health Care Quality Assurance*, *19*(6), 500–515.

Gordon, W. J., & Catalini, C. (2018). Gordon, W. J., and Catalini, C., 2018. Blockchain Technology for Healthcare: Facilitating the Transition to Patient-Driven Interoperability. Computational and






structural biotechnology journal 16: 224-230. *Computational and Structural Biotechnology Journal*, *16*, 224–230.

Griggs, K. N., Ossipova, O., Kohlios, C. P., Baccarini, A. N., Howson, E. A., & Hayajneh, T. (2018). Healthcare Blockchain System Using Smart Contracts for Secure Automated Remote Patient Monitoring. *Journal of Medical Systems*, *42*(7), 130.

Gropper, A. (2016). Powering the physician-patient relationship with HIE of one blockchain health IT. In *ONC/NIST Use of Blockchain for Healthcare and Research Workshop*. Gaithersburg, Maryland: ONC/NIST.

Hanington, B., & Martin, B. (2012). *Universal Methods of Design: 100 Ways to Research Complex Problems, Develop Innovative Ideas, and Design Effective Solutions*. Beverly, MA: Rockport Publishers.

Heston, T. (2017). *Why Blockchain Technology is Important for Healthcare Professionals*. Retrieved from https://papers.ssrn.com/abstract=3077455

Hofman, D., Lam, K., Shannon, C., Assadian, S., McManus, B., Ng, R., & Lemieux, V. L. (2018). Building Trust & Protecting Privacy: Analyzing Evidentiary Quality in a Blockchain Proof-of-Concept for Health Research Data Consent Management. *Proceedings of the IEEE Blockchain Conference,*. Presented at the Halifax, Canada. Halifax, Canada.

Horgan, R. P., & Kenny, L. C. (2011). 'Omic' technologies: Genomics, transcriptomics, proteomics and metabolomics. *The Obstetrician & Gynaecologist*, *13*(3), 189–195.

IBM. (2017). *Self-Sovereign Identity: Unraveling the terminology*. Retrieved from https://www.ibm.com/blogs/blockchain/2018/06/self-sovereign-identity-unraveling-the-terminology/





InterPARES. (2017). *InterPARES Trust Terminology Project: Key Blockchain Terms and Definitions*. Retrieved from http://arstweb.clayton.edu/interlex/blockchain/

Ivan, D. (2016, August). *Moving toward a blockchain-based method for the secure storage of patient records*. Presented at the ONC/NIST Use of Blockchain for Healthcare and Research Workshop, Gaithersburg, MA.

Joseph-Williams, N., Newcombe, R., Politi, M., Durand, M. A., Sivelli, S., & Stacey, D. (2014). Toward Minimum Standards for Certifying Patient Decision Aids: A Modified Delphi Consensus Process. *Medical Decision Making: An International Journal of the Society for Medical Decision Making*, *34*(6), 699–710.

Kaufman, D. J., Murphy-Bolinger, J., Scott, J., & Hudon, K. L. (2009). Public opinion about the importance of privacy in biobank research. *The American Journal of Human Genetics*, *85*(5), 643–654. https://doi.org/doi:10.1016/j.ajhg.2009.10.002

Kaur, H., Alam, M. A., Jameel, R., Mourya, A. K., & Chang, V. (2018). A Proposed Solution and Future Direction for Blockchain-Based Heterogeneous Medicare Data in Cloud Environment. *Journal of Medical Systems*, *42*(8), 156.

Krombholz, K., Judmayer, A., Gusenbauer, M., & Weippl, E. (2016, February). The other side of the coin: User experiences with bitcoin security and privacy. In *International conference on financial cryptography and data security* (pp. 555-580). Springer, Berlin, Heidelberg.

Kumar, V. (2012). *101 Design Methods: A Structured Approach for Driving Innovation in Your Organization*. Hoboken, N.J: Wiley.

LeRouge, C., & Wickramasinghe, N. (2013). A Review of User-Centered Design for Diabetes-Related Consumer Health Informatics Technologies. Journal of Diabetes Science and Technology. *Journal*






of *Diabetes Science and Technology*, *7*(4), 1039–1056. https://doi.org/doi.org/10.1177/193229681300700429

Li, X., Jiaing, R., Chen, T., Luo, X., & Wen, Q. (2017). A survey on the security of blockchain systems. *Archivx.Org*. Retrieved from https://arxiv.org/pdf/1802.06993.pdf

Linn, L. A., & Koo, M. B. (2016). *Blockchain for health data and its potential use in health it and health care related research*. Presented at the ONC/NIST Use of Blockchain for Healthcare and Research Workshop., Gaithersburg, MA.

Ljunggren, N. (2019). *Improving the usability of secure information storing within blockchain applications*. Retrieved from https://lup.lub.lu.se/student-papers/search/publication/8972293.

Mackey, T. K., & Nayyar, G. (2017). A review of existing and emerging digital technologies to combat the global trade in fake medicines. *Expert Opinion on Drug Safety*, *16*(5), 587–602.

Manzoni, C., Kia, D. A., Vandrovcova, J., Hardy, J., Wood, N., Lewis, N. W., & Ferrari, R. (2018). Genome, transcriptome and proteome: The rise of omics data and their integration in biomedical sciences. *Briefings in Bioinformatics*, *19*(2), 286–302. https://doi.org/doi:10.1093/bib/bbw114

McFerran, K. S., Heese, C., Medcalf, L., Murphy, M., & Fairchild, R. (2017). Integrating emotions into the critical interpretive synthesis. *Qualitative Health Research*, *27*(1), 13–23.

Medicalchain. (2018). *Whitepaper 2.0*. Retrieved from https://medicalchain.com/Medicalchain-Whitepaper-EN.pdf.

Morgan D.L (2005). *Focus Groups: Encyclopedia of social measurement*. Edited by: Kimberly K. New York: Elsevier, 51-57.

Neef, D. (2014). *Digital exhaust: What everyone should know about big data, digitization and digitally driven innovation*. New York, NY: Pearson Education.







Nissenbaum, H. (2010). *Privacy in context: Technology, policy, and the integrity of social life*. Stanford, CA: Stanford Law Books.

Nunnally, B., & Farkes, D. (2016). *UX Research: Practical Techniques for Designing Better Products*. New York, NY: O'Reilly Media, Inc.

Patel, V. (2018). A framework for secure and decentralized sharing of medical imaging data via blockchain consensus. *Health Informatics Journal*, 1–14. https://doi.org/doi/10.1177/1460458218769699

Peterson, K., Deedavuru, R., Kanjamata, P., & Boles, K. (2017). A Blockchain-Based Approach to Health Information Exchange Networks. *Proc. NIST Workshop Blockchain Healthcare*, *1*, 1–10.

Rosenberg, M. (2018, March 17). How Trump Consultants Exploited the Facebook Data of Thousands. *New York Times*. Retrieved from https://www.nytimes.com/2018/03/17/us/politics/cambridge-analytica-trump-campaign.html.

Sanderson, S. C., Linderman, M. D., Suckiel, S., Diaz, G. A., Zinberg, R. E., Ferryman, K., & Schadt, E. E. (2016). Motivations, concerns and preferences of personal genome sequencing research participants: Baseline findings from the HealthSeq project. *European Journal of Human Genetics : EJHG, 24*(1), 14–20. https://doi.org/doi:10.1038/ejhg.2015.118

Shabani, M., Bezuidenhout, L., & Borry, P. (2014). Shabani, M., Bezuidenhout, L., & Borry, P. (2014). Attitudes of research participants and the general public towards genomic data sharing: A systematic literature review. Expert review of molecular diagnostics, 14(8), 1053-1065. *Expert Review of Molecular Diagnostics, 14*(8), 10510653-.







Shi, X., & Wu, X. (2017). Shi, X. and Wu. X., 2017. An overview of human genetic privacy. Annals of the New York Academy of Sciences 1387 (1): 61-72. *Annals of the New York Academy of Sciences*, *1387*(1), 61–72.

Sovrin. (2019). *Governance Framework*. Retrieved from https://sovrin.org/wp-content/uploads/Sovrin-Governance-Framework-V2-Master-Document-V1.pdf

Stewart, J. M., Tittel, E., & Chapple, M. (2008). *Stewart, J. M., Tittel, E. and Chapple, M., 2008. CISSP: Certified information systems security professional study guide. New York: John Wiley & Sons.* New York, NY: John Wiley & Sons, Inc.

Swan, M. (2015). *Blockchain: Blueprint for a new economy.* New York, NY: O'Reilly Media, Inc.

Thorogood, A., & Zawati, M. H. (2015). Thorogood, A. and Zawati. M.H., 2015. International guidelines for privacy in genomic biobanking (or the unexpected virtue of pluralism). The Journal of Law, Medicine & Ethics 43 (4): 690-702. *The Journal of Law, Medicine & Ethics*, *43*(4), 690–702.

Tobin, A. (2018). *Sovrin: What goes on the ledger?* Retrieved from https://www.evernym.com/wp-content/uploads/2017/07/What-Goes-On-The-Ledger.pdf

Tobin, A., & Reed, D. (2017, March 28). *The Inevitable Rise of Self-Sovereign Identity*. Sovrin Foundation.

van Schaik, P., Jansen, J., Onibokun, J., Camp, J., & Kusey, P. (2018). Security and privacy in online social networking: Risk perceptions and precautionary behaviour. *Computers in Human Behavior*, *78*, 283–297. https://doi.org/doi:10.1016/j.chb.2017.10.007

Vaughan, L. K., & Srinivasasainagendra, V. (2013). Where in the genome are we? A cautionary tale of database use in genomics research. *Frontiers in Genetics*, *4*(38), 1–6.







W3C Community Group. (2019). *Decentralized Identifiers (DIDs) v0.11: Data Model and Syntaxes for Decentralized Identifiers (DIDs)*. Retrieved from https://w3c-ccg.github.io/did-spec/

Wharton, C., Rieman, J., Lewis, C., & Poison, P. (1994). The cognitive walkthrough method: A practitioner's guide. In *Usability Inspection Methods* (pp. 105–140). Retrieved from http://dl.acm.org/citation.cfm?id=189200.189214

Wittes, B. (2011). *Database: Digital Privacy and the Mosaic*. Retrieved from https://www.brookings.edu/research/databuse-digital-privacy-and-the-mosaic/

Xia, Q., Sifah, E. B., Amofa, S., & Zhang, X. (2017). BBDS: Blockchain-based data sharing for electronic medical records in cloud environments. *Information*, *8*(2), 44.

Xie, A., & Carayon, P. (2015). A systematic review of human factors and ergonomics (HFE)-based healthcare system redesign for quality of care and patient safety. *Ergonomics*, *58*(1), 33–49. https://doi.org/doi.org/10.1080/00140139.2014.959070

Young, K., & Vescent, H. (2018, August 28). 10 things you need to know about Self Sovereign Identity, part 1. Retrieved from He Paypers Insight into Payments website: https://www.thepaypers.com/expert-opinion/10-things-you-need-to-know-about-self-sovereign-identity-part-1/774556

Yue, L., Junqin, H., Shengzhi, Q., & Ruijin, W. (2017, August). *Yue, L., Junqin, H., Shengzhi, Q., and Ruijin, W. , 2017, August. Big data model of security sharing based on Blockchain. In 2017 3rd International Conference on Big Data Computing and Communications (BIGCOM) (pp. 117-121). IEEE.* 117–121. IEEE.







Zhang, P., White, J., Schmidt, D. C., & Lenz, G. (2017). *Applying software patterns to address interoperability in blockchain-based healthcare apps*. Retrieved from https://arxiv.org/pdf/1706.03700.pdf

Zhou, L., Wang, L., & Sun, Y. (2018). MIStore: A Blockchain-Based Medical Insurance Storage System. *Journal of Medical Systems*, *42*(8), 149.

Zhou, L., Wang, L., Sun, Y., & Lv, P. (2018). BeeKeeper: A Blockchain-based IoT System with Secure Storage and Homomorphic Computation. *IEEE Access*, *6*, 43472–43488.

Zyskind, G., Nathan, O., & Pentland, A. (2015). *Decentralizing privacy: Using blockchain to protect personal data*. MIT Media Lab.